\documentclass[reqno,centertags, 12pt]{amsart}
\usepackage{amsmath,amsthm,amscd,amssymb}
\usepackage{latexsym}
\newcommand{\bbD}{{\mathbb{D}}}

\newcommand{\bbC}{{\mathbb{C}}}

\newcommand{\calI}{{\mathcal I}}

\newcommand{\lb}{\label}
\newcommand{\f}{\frac}

\newcommand{\ti}{\tilde  }

\newcommand{\s}{\text{\rm{s}}}

\newcommand{\bi}{\bibitem}

\newcommand{\beq}{\begin{equation}}
\newcommand{\eeq}{\end{equation}}
\newcommand{\ba}{\begin{align}}
\newcommand{\ea}{\end{align}}
\newcommand{\veps}{\varepsilon}
\newcounter{smalllist}
\newenvironment{SL}{\begin{list}{{\rm\roman{smalllist})}}{%
\setlength{\topsep}{0mm}\setlength{\parsep}{0mm}\setlength{\itemsep}{0mm}%
\setlength{\labelwidth}{2em}\setlength{\leftmargin}{2em}\usecounter{smalllist}%
}}{\end{list}}

\DeclareMathOperator{\Real}{Re}

\allowdisplaybreaks
\numberwithin{equation}{section}
\newtheorem{theorem}{Theorem}[section]

\newtheorem*{p2.1}{Proposition 2.1}

\newtheorem{lemma}[theorem]{Lemma}

\theoremstyle{definition}

\newtheorem{conjecture}[theorem]{Conjecture}

\theoremstyle{remark}

\newcommand{\abs}[1]{\lvert#1\rvert}

\begin{document}
\title[Higher-Order Szeg\H{o} Theorems With Two Singular Points]
{Higher-Order Szeg\H{o} Theorems\\ With Two Singular Points}
\author[B. Simon and A. Zlato\v{s}]{Barry Simon$^{1}$ and Andrej Zlato\v{s}$^{2}$}

\thanks{$^1$ Mathematics 253-37, California Institute of Technology, Pasadena, CA 91125.
E-mail: bsimon@caltech.edu. Supported in part by NSF grant DMS-0140592}
\thanks{$^2$ Department of Mathematics, University of Wisconsin, Madison, WI 53706. E-mail:
andrej@math.wisc.edu}

\date{September 16, 2004}

\begin{abstract} We consider probability measures, $d\mu=w(\theta) \f{d\theta}{2\pi} +d\mu_\s$,
on the unit circle, $\partial\bbD$, with Verblunsky coefficients, $\{\alpha_j\}_{j=0}^\infty$.
We prove for $\theta_1\neq\theta_2$ in $[0,2\pi)$ and $(\delta\beta)_j=\beta_{j+1}$ that
\[
\int [1-\cos(\theta-\theta_1)][1-\cos(\theta-\theta_2)] \log w(\theta) \, \f{d\theta}{2\pi} >-\infty
\]
if and only if
\[
\sum_{j=0}^\infty \, \bigl|\bigl\{(\delta -e^{-i\theta_2}) (\delta
-e^{-i\theta_1}) \alpha\bigr\}_j\bigr|^2 +\abs{\alpha_j}^4 <\infty
\]
We also prove that
\[
\int (1-\cos\theta)^2 \log w(\theta)\, \f{d\theta}{2\pi} >-\infty
\]
if and only if
\[
\sum_{j=0}^\infty \abs{\alpha_{j+2}-2\alpha_{j+1} +\alpha_j}^2 + \abs{\alpha_j}^6 <\infty
\]
\end{abstract}

\maketitle

\section{Introduction} \lb{s1}

This paper is a contribution to the theory of orthogonal polynomials on the unit cirle (OPUC);
see \cite{GBk,OPUC1,OPUC2,Szb} for background. Throughout, $d\mu$ will be a nontrivial
probability measure on the unit circle, $\partial\bbD$, in $\bbC$, which we suppose has the
form
\begin{equation} \lb{1.1}
d\mu = w(\theta)\, \f{d\theta}{2\pi} + d\mu_\s
\end{equation}
where $d\mu_\s$ is singular with respect to Lebesgue measure $d\theta$ on $\partial\bbD$.

The Carath\'eodory and Schur functions, $F$ and $f$, associated to $d\mu$ are given for $z\in\bbD$ by
\begin{align}
F(z) &= \int \f{e^{i\theta}+z}{e^{i\theta}-z}\, d\mu(\theta) \lb{1.2} \\
&= \f{1+zf(z)}{1-zf(z)} \lb{1.3}
\end{align}
The Verblunsky coefficients $\{\alpha_j\}_{j=0}^\infty$ can be defined inductively
by the Schur algorithm
\begin{equation} \lb{1.4}
f(z) = \f{\alpha_0 + zf_1 (z)}{1+z\bar\alpha_0 f_1(z)}
\end{equation}
which defines $\alpha_0\in \bbD$ and $f_1$. Iterating gives
$\alpha_1, \alpha_2, \dots$ and $f_2, f_3, \dots$. That
$\alpha_j\in\bbD$ (rather than just $\bar\bbD$) follows from the
assumption that $d\mu$ is nontrivial, that is, has infinite
support so $f$ is not a finite Blaschke product. Actually,
\eqref{1.4} defines what are usually called Schur parameters; the
Verblunsky coefficients are defined by a recursion relation on the
orthogonal polynomials. The equality of these recursion
coefficients and the Schur parameters of \eqref{1.4} is a theorem
of Geronimus \cite{Ger44}; see \cite{OPUC1}. We will use the
definition in \eqref{1.4}.

The most famous result in OPUC is Szeg\H{o}'s theorem which, in Verblunsky's format \cite{V36},
says
\begin{equation} \lb{1.5}
\log \biggl(\, \prod_{j=0}^\infty (1-\abs{\alpha_j}^2)\biggr) =
\int \log (w(\theta)) \, \f{d\theta}{2\pi}
\end{equation}
In this expression, both sides are nonpositive (since
$\abs{\alpha_j}<1$, and Jensen's inequality implies $\int \log
(w(\theta))\, \f{d\theta}{2\pi}\leq \log (\int
w(\theta)\f{d\theta}{2\pi}) \leq \log (\mu(\partial\bbD))$).
Moreover, \eqref{1.5} includes the statement that both sides are
finite (resp.~$-\infty$) simultaneously. Thus \eqref{1.5} implies
a spectral theory result.

\begin{theorem}\lb{T1.1}
\begin{equation} \lb{1.6}
\int \log(w(\theta))\, \f{d\theta}{2\pi} > -\infty \Leftrightarrow \sum_{j=0}^\infty
\, \abs{\alpha_j}^2 <\infty
\end{equation}
\end{theorem}

This form of the theorem has caused considerable recent interest due to work of Deift-Killip
\cite{DeiftK} and Killip-Simon \cite{KS} which motivated a raft of papers
\cite{DenJDE,Kup1,Kup2,LNS,NPVY,Sim288,SZ,Zl2004}.

In \cite[Section~2.8]{OPUC1}, Simon found a higher-order analog to \eqref{1.6} that allows
$\log (w(\theta))$ to be singular at a single point:

\begin{theorem}\lb{T1.2}
\begin{equation} \lb{1.7}
\int (1-\cos\theta) \log(w(\theta))\, \f{d\theta}{2\pi} >-\infty
\Leftrightarrow \sum_{j=0}^\infty \, \abs{\alpha_{j+1} -\alpha_j}^2 +
\abs{\alpha_j}^4 <\infty
\end{equation}
\end{theorem}

This result allows a single singular point of order $1$ in $\log
(w(\theta))$  at $\theta=0$. By a simple rotation argument
\cite{OPUC1}, if $\cos(\theta)$ is replaced by
$\cos(\theta-\theta_1)$, $\abs{\alpha_{j+1}-\alpha_j}^2$ is
replaced by $\abs{\alpha_{j+1} -e^{-i\theta_1}\alpha_j}^2$.

Our goal in this paper is to analyze two singularities or a single double singularity.
We will prove that

\begin{theorem} \lb{T1.3} For $\theta_1\neq \theta_2$,
\begin{equation} \lb{1.8}
\begin{split}
\int (1-\cos&(\theta-\theta_1))(1-\cos(\theta-\theta_2))
\log(w(\theta)) \, \f{d\theta}{2\pi}
>-\infty \\
&\Leftrightarrow \sum_{j=0}^\infty \bigl|\bigr\{ \, (\delta
-e^{-i\theta_2})(\delta -e^{-i\theta_1}) \alpha\bigr\}_j\bigr|^2 +
\abs{\alpha_j}^4 <\infty
\end{split}
\end{equation}
\end{theorem}

In this theorem, $\delta$ is the operator on sequences
\begin{equation} \lb{1.9}
(\delta\alpha)_j =\alpha_{j+1}
\end{equation}
We will also prove a result for $\theta_1 =\theta_2$.

\begin{theorem}\lb{T1.4}
\begin{equation} \lb{1.10}
\int (1-\cos\theta)^2 \log (w(\theta)) \, \f{d\theta}{2\pi}
>-\infty \Leftrightarrow \sum_{j=0}^\infty \, \abs{\alpha_{j+2}
-2\alpha_{j+1} +\alpha_j}^2 + \abs{\alpha_j}^6 <\infty
\end{equation}
\end{theorem}

Again, one can replace $\cos(\theta)$ by $\cos(\theta-\theta_1)$
if $\{\alpha_{j+2} -2 \alpha_{j+1} +\alpha_j\}_j$ is replaced by
$\{(\delta-e^{-i\theta_1})^2\alpha\}_j$.

Given the form of these theorems, it is natural to conjecture the situation for arbitrarily many
singularities:

\begin{conjecture} \lb{Cn1.5} For $\{\theta_k\}_{k=1}^\ell$ distinct in $[0,2\pi)$,
\begin{equation} \lb{1.11}
\begin{split}
\int \prod_{k=1}^\ell &(1-\cos (\theta- \theta_k))^{m_k} \log
(w(\theta)) \, \f{d\theta}{2\pi}
>-\infty \\
&\Leftrightarrow \sum_{k=0}^\infty \biggl| \biggl\{\,
\prod_{k=1}^\ell [\delta - e^{-i\theta_k}]^{m_k} \alpha
\biggr\}_j\biggr|^2 + \abs{\alpha_j}^{2\max (m_k)+2}<\infty
\end{split}
\end{equation}
\end{conjecture}

Independently of our work, Denisov-Kupin \cite{DenKup2} have found conditions on the $\alpha$'s
equivalent to the left side of \eqref{1.11} being finite. However, their conditions are
complicated and even for the case $\sum_{k=1}^\ell m_k =2$, it is not clear they are
equivalent to the ones we have in Theorems~\ref{T1.3} and \ref{T1.4} (although they must be!).

In Section~\ref{s2}, we review the features we need of the relative Szeg\H{o} function which
will  play a critical role in our proofs, and we compute its first two Taylor coefficients.
In Section~\ref{s3}, we prove Theorem~\ref{T1.3} in the special case $\theta_1=0$, $\theta_2=\pi$,
and in Section~\ref{s4}, we prove Theorem~\ref{T1.4}. With these two warmups done, we turn to the
general result, Theorem~\ref{T1.3}, in Section~\ref{s5}. The details of this are sufficiently
messy that we do not think this direct approach is likely to yield our conjecture.

\smallskip
We would like to thank S.~Denisov for telling us of his work \cite{DenKup2}.

\smallskip
\section{The Relative Szeg\H{o} Function} \lb{s2}

In Section~2.9 of \cite{OPUC1}, Simon introduced the relative
Szeg\H{o} function, defined by
\begin{equation} \lb{2.1}
(\delta_0 D)(z) = \f{1-\bar\alpha_0 f(z)}{\rho_0} \, \f{1-zf_1(z)}{1-zf(z)}
\end{equation}
where
\begin{equation} \lb{2.2}
\rho_k = (1-\abs{\alpha_k}^2)^{1/2}
\end{equation}
and $f,f_1$ are given by \eqref{1.3} and \eqref{1.4}.

The key property of $\delta_0 D$ we will need and the reason it was introduced is

\begin{theorem}[\mbox{\cite[Theorem 2.9.3]{OPUC1}}] \lb{T2.1} Let $d\mu_1$ be the measure
whose Verblunsky coefficients are $(\alpha_1, \alpha_2, \dots)$. Let $w$ be given by
\eqref{1.1} and $w_1$ by
\begin{equation} \lb{2.3}
d\mu_1 = w_1(\theta) \, \f{d\theta}{2\pi} + d\mu_{1,\s}
\end{equation}
Suppose $w(\theta)\neq 0$ for a.e.~$e^{i\theta}$ in
$\partial\bbD$. Then the same is true for $w_1$ and
\begin{equation} \lb{2.4}
(\delta_0D)(z) =\exp\biggl( \f{1}{4\pi} \int \f{e^{i\theta}+z}{e^{i\theta}-z}\,
\log \biggl( \f{w(\theta)}{w_1(\theta)}\biggr)d\theta\biggr)
\end{equation}
\end{theorem}

As in \cite{KS,Sim288,SZ}, this is the basis for step-by-step sum rules, as we will see.

To prove Theorems \ref{T1.3} and \ref{T1.4}, we will need to start
with computing the first three Taylor coefficients of $\log
((\delta_0 D)(z))$.

\begin{theorem}\lb{T2.2} We have that
\begin{equation} \lb{2.5}
\log (\delta_0 D(z)) =A_0+A_1 z + A_2 z^2 +O(z^3)
\end{equation}
where
\begin{align}
A_0 &=\log \rho_0 \lb{2.6}  \\
A_1 &= \alpha_0 -\alpha_1 -\bar\alpha_0\alpha_1 \lb{2.7} \\
A_2 &= \tfrac12\, \alpha_0^2 -\tfrac12\, \alpha_1^2 + \alpha_1 -\alpha_2
-\alpha_1 \abs{\alpha_0}^2 + \alpha_2 \abs{\alpha_1}^2 -\bar\alpha_0 \alpha_2
\rho_1^2 + \tfrac12\, \bar\alpha_0^2 \alpha_1^2 \lb{2.8}
\end{align}
\end{theorem}

\begin{proof}  $f_2(0) = \alpha_2$, so
\[
f_1 = \f{zf_2+\alpha_1}{1+z\bar\alpha_1 f_2} = \alpha_1 + z\alpha_2 \rho_1^2 + O(z^2)
\]
Thus
\[
f= \f{zf_1 + \alpha_0}{1+z\bar\alpha_1 f_1} = \alpha_0 + z\alpha_1 \rho_0^2 +
z^2 \rho_0^2 (\alpha_2 \rho_1^2 -\bar\alpha_0 \alpha_1^2) + O(z^3)
\]
Plugging these into \eqref{2.1} yields the required Taylor coefficients.
\end{proof}

{\it Remarks.} 1. Denisov-Kupin \cite{DenKup2} do what is essentially the same
calculation using the CMV matrix.

\smallskip
2. \eqref{3.2} and \eqref{3.3} below show that \eqref{2.4} implies
\begin{align}
\int \log \biggl( \f{w(\theta)}{w_1(\theta)}\biggr) \f{d\theta}{2\pi} &= 2A_0 \lb{2.9}  \\
\int \log \biggl( \f{w(\theta)}{w_1(\theta)}\biggr)e^{-im\theta}\,
\f{d\theta}{2\pi}
&= \begin{cases} A_m & m=1,2 \\
\bar A_{-m} & m=-1,-2
\end{cases} \lb{2.10}
\end{align}

\smallskip

\section{Singularities at Antipodal Points} \lb{s3}

As a warmup, in this section we prove the following, which is
Theorem~\ref{T1.3} for $\theta_1 =0$, $\theta_2=\pi$. By the
remark after Theorem \ref{T1.2} this also gives the result for any
antipodal $\theta_1$ and $\theta_2$.

\begin{theorem}\lb{T3.1}
\begin{equation} \lb{3.1}
\int (1-\cos^2 (\theta)) \log w(\theta)\, \f{d\theta}{2\pi} >-\infty
\Leftrightarrow \sum_{j=0}^\infty \, \abs{\alpha_{j+2} -\alpha_j}^2 + \abs{\alpha_j}^4 <\infty
\end{equation}
\end{theorem}

{\it Remark.}  Let $\alpha_j$ be given and let $\beta_j$ be the
sequence $(\alpha_0, 0, \alpha_1, 0, \alpha_2, 0, \dots)$. Then
(see Example~1.6.14 of \cite{OPUC1}), $w^{(\beta)} (\theta) =\f12
w^{(\alpha)}(\f12\theta)$ and the RHS of \eqref{3.1} for $\beta =$
the RHS of \eqref{1.7} for $\alpha$. Thus \eqref{3.1} for $\beta$
is \eqref{1.7} for $\alpha$. This shows, in particular, that if a
result like \eqref{3.1} holds, it must involve $\abs{\alpha_j}^4$,
rather than, say, $\abs{\alpha_j}^6$.

\smallskip
We begin by noting that if $Q(\theta)$ is real and
\begin{equation} \lb{3.2}
Q(\theta) =\sum_{n=-\infty}^\infty b_n e^{in\theta}
\end{equation}
then
\begin{equation} \lb{3.3}
\int \f{e^{i\theta}+z}{e^{i\theta}-z}\, Q(\theta) \f{d\theta}{2\pi} =
b_0 + 2\sum_{n=1}^\infty b_n z^n
\end{equation}
since $(e^{i\theta}+z)/(e^{i\theta}-z) = 1+2\sum_{n=1}^\infty z^n
e^{-in\theta}$. Thus, by \eqref{2.9}, \eqref{2.10}, and
\begin{equation} \lb{3.4}
1-\cos^2 (\theta) =\tfrac14\, (2-e^{2i\theta} -e^{-2i\theta})
\end{equation}
we have
\begin{equation} \lb{3.5}
\int (1-\cos^2(\theta))\log \biggl( \f{w(\theta)}{w_1(\theta)}\biggr) \f{d\theta}{2\pi}
=A_0 -\tfrac12\, \Real (A_2)
\end{equation}
with $A_0$ given by \eqref{2.6} and $A_2$ by \eqref{2.8}.

\begin{lemma} \lb{L3.2} We have that
\begin{equation} \lb{3.6}
A_0 -\tfrac12\, \Real (A_2) =B_0 + C_0 + D_0 + F_0 -F_1 + G_0 -G_2
\end{equation}
where
\begin{align}
B_j &= \tfrac12\, \bigl[ \log (1-\abs{\alpha_j}^2) +
\abs{\alpha_j}^2 +
\tfrac 12\abs{\alpha_j}^4 \bigr] \lb{3.7} \\
C_j &= -\tfrac14\, (1-\abs{\alpha_{j+1}}^2) \abs{\alpha_j-\alpha_{j+2}}^2 \lb{3.8} \\
D_j &= -\tfrac18\, (\abs{\alpha_{j+1}^2 + \alpha_j^2}^2 + 4\abs{\alpha_j \alpha_{j+1}}^2)\lb{3.9} \\
F_j &= -\tfrac12\, \Real ( \tfrac12\, \alpha_j^2 + \alpha_{j+1} -\alpha_{j+1} \abs{\alpha}^2)
+ \tfrac14\, \abs{\alpha_{j+1}}^2 \abs{\alpha_j}^2 - \tfrac18\, \abs{\alpha_j}^4 \lb{3.10} \\
G_j &= -\tfrac14\, \abs{\alpha_j}^2 \notag
\end{align}
\end{lemma}

{\it Remark.} \eqref{3.5}/\eqref{3.6} is thus the step-by-step sum rule in the spirit of
\cite{KS,Sim288,SZ}.

\begin{proof} This is a straightforward but tedious calculation. The first term
in $B_0$ is just $A_0$ (since $\log \rho_j =\f12 \log (1-\abs{\alpha_j}^2)$).
$A_2$ is responsible for the $\Real(\cdot)$ terms in $F_0 - F_1$
and the cross-terms in $\abs{\alpha_j-\alpha_{j+2}}^2$ and
$\abs{\alpha_{j+1}^2 +\alpha_j^2}^2$. The $\abs{\alpha_j}^2 +
\abs{\alpha_{j+2}}^2$ term in $C_0$ is turned into
$2\abs{\alpha_j}^2$ by $G_0 -G_{2}$, and then cancelled by the
$\abs{\alpha_j}^2$ term in $B_0$. Similarly, the $\abs{\alpha_j}^4
+ \abs{\alpha_{j+1}}^4$ in $D_0$ (after adding the
$\abs{\alpha_j}^4$ terms in $F_0-F_1$) cancels the
$\abs{\alpha_j}^4$ term in $B_0$. Finally, the
$\abs{\alpha_{j+1}}^2 (\abs{\alpha_j}^2 + \abs{\alpha_{j+2}}^2)$
term in $C_0$ (after being turned into
$2\abs{\alpha_{j+1}}^2\abs{\alpha_{j}}^2$ by the
$\abs{\alpha_{j+1}}^2 \abs{\alpha_j}^2$ term in $F_0-F_1$) cancels
the $4\abs{\alpha_j\alpha_{j+1}}^2$ term in $D_0$.
\end{proof}

By iterating \eqref{3.5}/\eqref{3.6} and noting the cancellations from the telescoping
$F_j -F_{j+1}$ and $G_j -G_{j+2}$ yields
\begin{equation} \lb{3.11}
\begin{split}
\int (1- &\cos^2 (\theta)) \log \biggl( \f{w(\theta)}{w_{2m}(\theta)}\biggr) \f{d\theta}{2\pi} \\
&= F_0 -F_{2m} + G_0 + G_1 - G_{2m} -G_{2m+1} + \sum_{j=0}^{2m-1}
(B_j + C_j + D_j)
\end{split}
\end{equation}

As a final preliminary, we need,
\begin{lemma}\lb{L3.3}
\begin{SL}
\item[{\rm{(i)}}] $\abs{F_j} \leq \f{13}8$; $\abs{G_j} \leq \f14$
%\item[{\rm{(ii)}}] $0\geq B_j \geq - \abs{\alpha_j}^2/6$
\item[{\rm{(ii)}}] $\abs{\alpha_j} <\f12 \Rightarrow
c_1\abs{\alpha_j}^6 \le -B_j \le c_2\abs{\alpha_j}^6$ for some
$c_2>c_1>0$. \item[{\rm{(iii)}}] $\abs{\alpha_{j+1}}^4 +
\abs{\alpha_j}^4 \leq -8D_j \leq 4 (\abs{\alpha_{j+1}}^4
+\abs{\alpha_j}^4)$
\end{SL}
\end{lemma}

\begin{proof} (i) follows from $\abs{\alpha_j}\leq 1$, (ii) from $-\log (1-x) =
\sum_{j=1}^\infty x^j/j$, and (iii) by noting that $2\Real
(\alpha_j^2 \alpha_{j+1}^2) + 2\abs{\alpha_j^2 \alpha_{j+1}^2}
\geq 0$ and repeated use of $\abs{xy} \leq \f12 (|x|^2 + |y|^2)$.
\end{proof}

%{\it Remark.} (ii) is actually not needed!

\begin{proof}[Proof of Theorem~\ref{T3.1}] We follow the strategy of \cite{KS} as modified by
\cite{SZ}. Suppose first that the RHS of \eqref{3.1} holds. Let
$w^{(n)}$ be the weight for the $n^{\rm th}$ Bernstein-Szeg\H{o}
approximation with Verblunsky coefficients $(\alpha_0, \alpha_1,
\dots, \alpha_{n-1}, 0, \dots, 0, \dots)$. By \eqref{3.11} and
$(w^{(n)})_{2m}\equiv 1$ for large $m$,
\[
\int (1-\cos^2 (\theta)) \log (w^{(n)}(\theta))\, \f{d\theta}{2\pi} =
F_0^{(n)} + G_0^{(n)} + G_1^{(n)} + \sum_{j=0}^{n-1} (B_j^{(n)} + C_j^{(n)} + D_j^{(n)})
\]
so, by Lemma~\ref{L3.3}, $\abs{\alpha_j}^6 \leq
\abs{\alpha_j}^4\to 0$, and RHS of \eqref{3.1},
\begin{equation} \lb{3.12}
\inf_n \biggl[ \int (1-\cos^2 (\theta)) \log (w^{(n)} (\theta))
\f{d\theta}{2\pi} \biggr] > -\infty
\end{equation}
Up to a constant, $\int (1-\cos^2 (\theta)) \log
w(\theta)\f{d\theta}{2\pi}$ is an entropy and so upper
semicontinuous \cite{KS}. Thus \eqref{3.12} implies
\begin{equation} \lb{3.13}
\int (1-\cos^2 (\theta)) \log w(\theta)\, \f{d\theta}{2\pi} >
-\infty
\end{equation}

Conversely, suppose \eqref{3.13} holds. Since $\int
(1-\cos^2(\theta))\log (w_{2m}(\theta)) \f{d\theta}{2\pi}$ is an
entropy up to a constant, it is bounded above \cite{KS}, and so
the left side of \eqref{3.11} is bounded below as $m$ varies.

Since $F$ and $G$ are bounded and $B,C,D$ are negative, we
conclude
\[
\sum_{j=0}^\infty -(B_j +C_j +D_j) <\infty
\]
Since $\sum (-D_j) <\infty$, Lemma~\ref{L3.3} implies $\sum
\abs{\alpha_j}^4 <\infty$. This implies $\alpha_j \to 0$, so $\sum
(-C_j) <\infty$ implies $\sum \abs{\alpha_j -\alpha_{j+2}}^2
<\infty$.
\end{proof}

\smallskip

\section{Singularity of Order $2$} \lb{s4}

Our goal here is to prove Theorem~\ref{T1.4}. Since
\begin{align*}
(1-\cos\theta)^2 &= \tfrac14\, (2-e^{i\theta}-e^{-i\theta})^2 \\
&= \tfrac32 - e^{i\theta} -e^{-i\theta} + \tfrac14 \, e^{2i\theta} + \tfrac14\, e^{-2i\theta}
\end{align*}
we see, by \eqref{2.9}/\eqref{2.10} that
\begin{equation} \lb{4.1}
\int \log \biggl( \f{w(\theta)}{w_1(\theta)}\biggr) (1-\cos\theta)^2 \, \f{d\theta}{2\pi} =
3A_0 -2\Real (A_1) + \tfrac12\, \Real (A_2)
\end{equation}
with $A_0, A_1, A_2$ given by \eqref{2.6}--\eqref{2.8}.

\begin{lemma}\lb{L4.1} The RHS of \eqref{4.1} $= H_0 + I_0 +J_0 + K_0 -K_1 + L_0 -L_2$ where
\begin{align*}
H_j &=\tfrac32\, [\log (1-\abs{\alpha_j}^2) + \abs{\alpha_j}^2] \\
I_j &= -\tfrac14\, \abs{\alpha_{j+2} -2\alpha_{j+1} +\alpha_j}^2 \\
J_j &=\tfrac14\, (\alpha_j \bar\alpha_{j+2} + \bar\alpha_j \alpha_{j+2})
\abs{\alpha_{j+1}}^2 + \tfrac18\, (\alpha_j^2 \bar\alpha_{j+1}^2 + \bar\alpha_j^2 \alpha_{j+1}) \\
K_j &= -2\Real (\alpha_j) + \tfrac14\, \Real (\alpha_j^2)  \\
&\qquad +\tfrac12\, \Real (\alpha_{j+1}) - \tfrac12\, \Real
(\alpha_{j+1} \abs{\alpha_j}^2) + \Real [\bar\alpha_{j+1}
\alpha_j] -
\abs{\alpha_j}^2 \\
L_j &= -\tfrac14\, \abs{\alpha_j}^2
\end{align*}
\end{lemma}

\begin{proof} The non-cross-terms in $I_0$ are
\[
-\tfrac14\, (\abs{\alpha_2}^2 + 4\abs{\alpha_1}^2 + \abs{\alpha_0}^2) =
-\tfrac32\, \abs{\alpha_0}^2 + (\abs{\alpha_0}^2 - \abs{\alpha_1}^2) +
\tfrac14\, (\abs{\alpha_0}^2 -\abs{\alpha_2}^2)
\]
which cancel the $\abs{\alpha_0}^2$ term in $H_0$, the final
$\abs{\alpha_j}^2$ term in $K_0-K_1$, and the $L_0-L_2$ term.

The cross-terms in $I_0$ are
\[
\begin{split}
-\tfrac12\, \Real (\bar\alpha_2 &\alpha_0) - \Real (\bar\alpha_2 \alpha_1 + \bar\alpha_1 \alpha_0) \\
&=-\tfrac12\, \Real (\bar\alpha_2\alpha_0) + 2\Real (\bar\alpha_0\alpha_1) -
\Real (\bar\alpha_0\alpha_1) + \Real (\bar\alpha_1\alpha_2)
\end{split}
\]
The first term comes from the piece of $\tfrac 12\Real (A_2)$
(since $\bar\alpha_0\alpha_2\rho_1^2 = \bar\alpha_0 \alpha_2
(1-\abs{\alpha_1}^2)$, the second from the last term in $-2\Real
(A_1)$, and the last two are cancelled by the $\Real
(\bar\alpha_{j+1}\alpha_j)$ term in $K_0 -K_1$.

The $\alpha_0-\alpha_1$ term in $A_0$ leads to the first term in
$K_1 -K_0$. The first term in $J_0$ comes from the second half of
$\bar\alpha_0\alpha_2\rho_1^2 =\bar\alpha_0\alpha_2 -\bar\alpha_0
\alpha_2 \abs{\alpha_1}^2$ (the first half in this expression gave
a cross-term in $I_j$). The second term in $J_0$ is the $\f12
\bar\alpha_0^2 \alpha_1^2$ term in $A_2$.

The remaining terms in $A_2$ give precisely the remaining terms in
$K_0 -K_1$.
\end{proof}

\begin{lemma} \lb{L4.2} The RHS of \eqref{4.1} $= \ti H_0 + \ti I_0 + \ti J_0 + \ti K_0 -
\ti K_1 + \ti L_0 - \ti L_2$, where
\begin{align*}
\ti H_j &= \tfrac32\, \bigl[\log (1-\abs{\alpha_j}^2) +
\abs{\alpha_j}^2 +
\tfrac 12\abs{\alpha_j}^4 \bigr] \\
\ti I_j &= I_j \\
\ti J_j &= -\tfrac14\, \abs{\alpha_{j+1}}^2 \abs{\alpha_j -\alpha_{j+2}}^2 - \tfrac18\,
\abs{\alpha_{j+1}^2 -\alpha_j^2}^2 - \tfrac14\, (\abs{\alpha_{j+1}}^2 - \abs{\alpha_j}^2)^2\\
\ti K_j &= K_j -\tfrac38\, \abs{\alpha_j}^2 - \tfrac14\, \abs{\alpha_{j+1}}^2 \abs{\alpha_j}^2 \\
\ti L_j &= L_j
\end{align*}
\end{lemma}

\begin{proof} The non-cross-terms in the last two terms in $\ti J_0$ give
\[
-\tfrac38\, (\abs{\alpha_0}^4 + \abs{\alpha_1}^4) = -\tfrac34\,
\abs{\alpha_0}^4 + \tfrac38 (\abs{\alpha_0}^4 -\abs{\alpha_1}^4)
\]
The first term cancels the $\ti H_0 - H_0$ term, and the second, the first term in
$(K_0 - \ti K_0) - (K_1 -\ti K_1)$.

The cross-term in $-\f14 (\abs{\alpha_{j+1}}^2 -
\abs{\alpha_j}^2)^2$ and the non-cross-terms in $-\f14
\abs{\alpha_{j+1}}^2 \abs{\alpha_j -\alpha_{j+2}}^2$ combine to
$-\f14 \abs{\alpha_{j+2}}^2 \abs{\alpha_{j+1}}^2 + \tfrac14
\abs{\alpha_{j+1}}^2 \abs{\alpha_j}^2$ and are cancelled by the
second term in $(K_0 -\ti K_0) - (K_1 -\ti K_1)$. The cross-term
in $-\f18 \abs{\alpha_{j+1}^2 -\alpha_j^2}^2$ is the second term
in $J_0$ and finally, the cross-term in $-\f14
\abs{\alpha_{j+1}}^2 \abs{\alpha_j -\alpha_{j+2}}^2$ is the first
term in $J_0$.
\end{proof}

\begin{lemma}\lb{L4.3}
\begin{SL}
\item[{\rm{(i)}}] $\abs{\ti K_j} \leq \f{47}{8}$; $\abs{\ti L_j}
\le \f14$
%\item[{\rm{(ii)}}] $0\geq \ti H_j \geq - \abs{\alpha_j}^6/2$
\item[{\rm{(ii)}}] $\abs{\alpha_j} < \f12 \Rightarrow
d_1\abs{\alpha_j}^6 \le -\ti H_j \le d_2\abs{\alpha_j}^6$ for some
$d_2>d_1>0$. \item[{\rm{(iii)}}] $\ti J_j \le 0$
\item[{\rm{(iv)}}] $\sum_{j=0}^\infty (-\ti I_j) +
\abs{\alpha_j}^6 <\infty \Rightarrow \sum_{j=0}^\infty
\abs{\alpha_{j+1} -\alpha_j}^3 <\infty$ \item[{\rm{(v)}}]
$\sum_{j=0}^\infty (-\ti I_j) + \abs{\alpha_j}^6 <\infty
\Rightarrow \sum_{j=0}^\infty (-\ti J_j) <\infty$
\end{SL}
\end{lemma}

{\it Remark.} (iv) is essentially a discrete version of the
inequality of Gagliardo \cite{Gag} and Nirenberg \cite{Nir}.

\begin{proof} (i) follows from $\abs{\alpha_j} <1$, (ii) is just (ii)
of Lemma~\ref{L3.3} (since $\ti H_j =3B_j$), and (iii) is trivial.

To prove (iv), we let $\delta$ be given by \eqref{1.9} and let
\begin{equation} \lb{4.2}
\partial = \delta -1
\end{equation}
so since $\delta^* =\delta^{-1}$ ($\delta$ is unitary on $\ell^2$), we have
\begin{equation} \lb{4.3}
\partial^* = \delta^* -1 = -\delta^{-1} \partial = -\delta^*\partial
\end{equation}

As a result, if $\alpha$ is a finite sequence, then
\begin{align}
\sum_n \, \abs{(\partial\alpha)_n}^3&= \sum_n (\partial\alpha)_n
(\partial\bar\alpha)_n
\abs{\partial\alpha}_n \notag \\
&=-\sum_n (\delta\alpha)_n [\partial \{(\partial\bar\alpha)
\abs{\partial\alpha}\}]_n \lb{4.4}
\end{align}
Moreover, we have a discrete Leibnitz rule,
\begin{align}
\partial (fg) &= (\delta f)(\delta g)-fg \notag \\
&= (\delta f) \partial g + (\partial f)g \lb{4.5}
\end{align}
and since $\abs{a-b} \geq \abs{a}-\abs{b}$ by the triangle inequality,
\begin{equation} \lb{4.6}
\abs{\partial\abs{f}} \leq \abs{\partial f}
\end{equation}
which is a discrete Kato inequality.

By \eqref{4.5},
\[
\partial \{(\partial \bar\alpha)\abs{\partial\alpha}\} = [\delta(\partial\bar\alpha)]
\partial \abs{\partial\alpha} + (\partial^2 \bar\alpha)\abs{\partial\alpha}
\]
so, by \eqref{4.6},
\[
\abs{\partial \{(\partial\bar\alpha)\abs{\partial\alpha}\}} \leq
\abs{\partial^2 \alpha} \, \abs{\delta(\partial\bar\alpha)} +
\abs{\partial^2 \alpha} \, \abs{\partial\alpha}
\]
Using H\"older's inequality with $\f16 + \f12 + \f13 =1$ and
\eqref{4.4}, we get
\[
\|\partial\alpha\|_3^3 \leq 2 \| \alpha\|_6 \|\partial^2 \alpha\|_2 \|\partial\alpha\|_3
\]
(because $\|\delta\alpha\|_p=\|\alpha\|_p$), so
\begin{equation} \lb{4.7}
\sum_n \, \abs{(\partial\alpha)_n}^3 \leq 2^{3/2} \biggl(\,\sum_n \, \abs{\alpha_n}^6\biggr)^{1/4}
\biggl(\, \sum_n \, \abs{(\partial^2 \alpha)_n}^2\biggr)^{3/4}
\end{equation}
Having proven \eqref{4.7} for $\alpha$'s of finite support, we get
it for any $\alpha$ with the right side finite since $\sum_n
\abs{\alpha_n}^6 <\infty$ implies $\alpha_n\to 0$, which allows
one to cut off $\alpha$ at $N$ and take $N\to\infty$ in
\eqref{4.7}. But \eqref{4.7} implies (iv).

To prove (v), we control the individual terms in $\sum (-\ti
J_j)$. First,
\begin{align*}
\|\abs{\alpha}^2 \abs{\delta^2 \alpha -\alpha}^2\|_1 &\leq \| \,
\alpha^2\|_3 \,
\| \abs{\delta^2 \alpha -\alpha}^2\|_{3/2} \\
\intertext{(by H\"older's inequality with $\f13 + \f23 =1$)} &\leq
4 \|\alpha\|_6^2\, \|\partial\alpha\|_3^2 <\infty
\end{align*}
(by first using $\|\delta^2\alpha-\alpha\|_3\le
2\|\partial\alpha\|_3$ and then (iv)). Next,
\[
\abs{\alpha_{j+1}^2 -\alpha_j^2}^2 \leq (\abs{\alpha_{j+1}} + \abs{\alpha_{j+1}})^2
\abs{\alpha_{j+1} -\alpha_j}^2
\]
can be controlled as the first term was and the final term is
controlled in the same way since $|\alpha_{j+1}|^2-|\alpha_j|^2
\le |\alpha_{j+1}^2-\alpha_j^2|$.
\end{proof}

\begin{proof}[Proof of Theorem~\ref{T1.4}] Suppose first that the right-hand side of
\eqref{1.10} holds, that is, $\alpha\in\ell^6$ and
$\partial^2\alpha\in\ell^2$. Iterate $n$ times
\eqref{4.1}/Lemma~\ref{L4.2} for the $n^{\rm th}$
Bernstein-Szeg\H{o} approximation (with weight $w^{(n)}$) to
obtain
\[
\inf_n\, \biggl[ \int (1-\cos\theta)^2 \log (w^{(n)}(\theta))
\,\f{d\theta}{2\pi} \biggr] > -\infty
\]
since the left side is just
\[
\inf_n \Big[\ti K_0^{(n)}+\ti L_0^{(n)}+\ti L_1^{(n)}
+\sum_{j=0}^{n-1} (\ti H_j^{(n)} + \ti I_j^{(n)} + \ti J_j^{(n)})
\Big]
\]
which is finite by Lemma~\ref{L4.3} and the hypothesis. Again we
have that $\int (1-\cos\theta)^2 \log w(\theta) \f{d\theta}{2\pi}$
is an entropy up to a constant and so upper semicontinuous. Thus
RHS of \eqref{1.10} $\Rightarrow$ LHS of \eqref{1.10}.

For the opposite direction, as in the last section, we use
iterated \eqref{4.1}/Lemma~\ref{L4.2} plus the fact that $\int
(1-\cos\theta)^2 \log (w_{2m}(\theta)) \f{d\theta}{2\pi}$ is
bounded from above to conclude
\[
\sum_{j=0}^\infty -(\ti H_j + \ti I_j + \ti J_j) <\infty
\]
Since each is positive, $\sum (-\ti H_j) <\infty$, which implies
$\sum \abs{\alpha_j}^6 <\infty$ by (ii) of Lemma~\ref{L4.3}, and
$\sum_{j=0}^\infty (-\ti I_j)<\infty$, which implies
$\partial^2\alpha\in\ell^2$.
\end{proof}

\smallskip
\section{The General Case} \lb{s5}

Finally, we turn to the general case of Theorem \ref{T1.3}, and we
define
\begin{equation} \lb{5.1}
\calI_m \equiv \int \big[1-\cos(\theta-\theta_1)\big]
\big[1-\cos(\theta-\theta_2)\big] \log \biggl(
\f{w(\theta)}{w_m(\theta)}\biggr) \f{d\theta}{2\pi}
\end{equation}
%for $\theta_2-\theta_1\neq 0,\pi$.
Using \eqref{2.9} and \eqref{2.10}, we obtain
\begin{equation} \lb{5.2}
\calI_1  = \frac
{4+e^{i(\theta_1-\theta_2)}+e^{-i(\theta_1-\theta_2)}} 4 A_0 -
\Real \big[ (e^{i\theta_1}+e^{i\theta_2}) A_1 \big] + \tfrac 12
\Real \big[ e^{i(\theta_1+\theta_2)} A_2 \big]
\end{equation}

The situation is now somewhat more complicated than in the
previous sections and it will be more convenient to work with
$\calI_m$ from the start, only keeping track of the essential
components of the sums (analogs of $\sum (B_j+C_j+D_j)$ and
$\sum (\ti H_j+\ti I_j+\ti J_j)$ above) and ignore the ones that
are always bounded and hence irrelevant for us (analogs of
$F_0-F_1+G_0+G_1-G_m-G_{m+1}$ and $\ti K_0-\ti K_m+\ti L_0+\ti
L_1-\ti L_m+\ti L_{m+1}$). Hence substituting
\eqref{2.6}--\eqref{2.8} in \eqref{5.2} and iterating, we obtain
\begin{align*}
\calI_m & = C_{\alpha,m} + \frac
{4+e^{i(\theta_1-\theta_2)}+e^{-i(\theta_1-\theta_2)}}4
\sum_{j=0}^{m-1} \log (1-|\alpha_j|^2)
\\ & + \sum_{j=0}^{m-1} \Real \Big\{
\big( e^{i\theta_1}+e^{i\theta_2} \big) \alpha_{j+1}\bar\alpha_j -
\tfrac 12 e^{i(\theta_1+\theta_2)} \big[
\alpha_{j+2}\bar\alpha_j(1-|\alpha_{j+1}|^2) -\tfrac 12
\alpha_{j+1}^2\bar\alpha_j^2 \big] \Big\}
\end{align*}
where
\begin{align*}
C_{\alpha,m} & \equiv -\Real \big[ (e^{i\theta_1}+e^{i\theta_2})
(\alpha_0-\alpha_m) \big]
\\ &+ \tfrac 12 \Real \big[
e^{i(\theta_1+\theta_2)} \big(\tfrac 12 \alpha_0^2 - \tfrac 12
\alpha_m^2 + \alpha_1 - \alpha_{m+1} - \alpha_1|\alpha_0^2| +
\alpha_{m+1}|\alpha_m|^2 \big) \big]
\end{align*}
We let
\[
\beta_j\equiv \alpha_j e^{i(\theta_1+\theta_2)j/2}
\]
and
\[
a\equiv \tfrac 12 \big(
e^{i(\theta_1-\theta_2)/2}+e^{-i(\theta_1-\theta_2)/2} \big)
\in(-1,1)
\]
We will assume $a\neq 0$ since the case when $\theta_1$ and
$\theta_2$ are antipodal follows from Theorem \ref{T3.1}. With
$C_{\beta,m}\equiv C_{\alpha,m}$ and all the sums taken from $0$
to $m-1$, the above becomes
\begin{align}
\calI_m  = & C_{\beta,m} + \big(\tfrac 12 +a^2 \big) \sum \log
(1-|\beta_j|^2)  + a \sum \big[ \beta_{j+1}\bar\beta_j +
\bar\beta_{j+1}\beta_j \big] \notag
\\ & -\tfrac 14 \sum \big[ \beta_{j+2}\bar\beta_j (1-|\beta_{j+1}|^2)
+ \bar\beta_{j+2}\beta_j (1-|\beta_{j+1}|^2) \big] \notag
\\ & + \tfrac 18 \sum \big[ \beta_{j+1}^2\bar\beta_j^2 + \bar\beta_{j+1}^2\beta_j^2
\big] \lb{5.3}
\end{align}
In the following manipulations with the sums, we will use
$C_{\beta,m}$ as a general pool/depository of terms that will be
added/left over in order to keep all the sums from $0$ to $m-1$.
Its value will therefore change along the argument, but it will
always depend on a few $\beta_j$'s with $j$ close to $0$ or $m$ only
(i.e., it will gather all the ``irrelevant'' terms) and will
always be bounded by a universal constant.

\begin{lemma} \lb{L5.1}
With $C_{\beta,m}$ universally bounded, we have
\begin{align}
\calI_m =& C_{\beta,m} + \big(\tfrac 12 +a^2 \big) \sum \big[\log
(1-|\beta_j|^2) + |\beta_j|^2 + \tfrac 12 |\beta_j|^4 \big] \notag
\\ & - \tfrac 14 \sum (1-|\beta_{j+1}|^2)
\big|\beta_{j+2}-2a\beta_{j+1}+\beta_j \big|^2 \notag
\\ & -\tfrac 14 \sum
|\beta_{j+1}|^2 \big|\beta_{j+2}-2a\beta_{j+1} \big|^2 -\tfrac 14
\sum |\beta_{j+1}|^2 \big| \beta_{j}-2a\beta_{j+1} \big|^2 \notag
\\ & - \tfrac 18 \sum \big|\beta_{j+1}^2-\beta_j^2 \big|^2 +
\tfrac{1}2 a^2 \sum|\beta_j|^4 \lb{5.4}
\end{align}
with all the sums taken from $0$ to $m-1$.
\end{lemma}

{\it Remarks.} 1. This enables us to prove the ``$\Leftarrow$''
part of \eqref{1.8} (even if $a=0$) since
\begin{equation} \lb{5.5}
\big| \big\{ (\delta-e^{-i\theta_2})(\delta-e^{-i\theta_1}) \alpha
 \big\}_j \big| = \big|\beta_{j+2}-2a\beta_{j+1}+\beta_j \big|
\end{equation}
But to prove the other implication, we first need to deal with the
last sum in \eqref{5.4}, which has the ``wrong'' sign.

\smallskip
2. Note that we actually did not need to exclude the case $a=0$
since then the last sum in \eqref{5.4} vanishes and an examination
of \eqref{5.4} shows that $\lim_{m\to\infty} \calI_m>-\infty$ if
and only if the RHS of \eqref{1.8} holds. An argument from the
proofs of Theorems \ref{T1.3} and \ref{T1.4} then gives the
``$\Rightarrow$'' part of \eqref{1.8}.

\begin{proof}
Multiplying out the terms in the second, third, and fourth sums of
\eqref{5.4} and after obvious cancellations, we are left with
\[
-\tfrac 14 \sum \Big[ |\beta_{j+1}|^2 \big( 4a^2 |\beta_{j+1}|^2 -
\beta_{j+2}\bar\beta_j -\bar\beta_{j+2}\beta_j \big) +
\big|\beta_{j+2}-2a\beta_{j+1}+\beta_j \big|^2 \Big]
\]
But this is just
\begin{equation} \lb{5.6}
-\tfrac 14\sum \big[ |\beta_{j+2}|^2 + 4a^2|\beta_{j+1}|^2 +
|\beta_{j}|^2 + 4a^2 |\beta_{j+1}|^4 \big]
\end{equation}
plus the second and third sums in \eqref{5.3}, the latter written
as
\[
\tfrac 12 a \sum [ \beta_{j+2}\bar\beta_{j+1} +
\bar\beta_{j+2}\beta_{j+1} +\beta_{j+1}\bar\beta_j +
\bar\beta_{j+1}\beta_j ]
\]
(with $C_{\beta,m}$ keeping the change). Adding the fifth and
sixth sums in \eqref{5.4} to \eqref{5.6} and subtracting the last
sum in \eqref{5.3}, we obtain
\[
-\tfrac 14 \sum (2+4a^2)|\beta_j|^2 - \tfrac 18 \sum
(2+4a^2)|\beta_j|^4
\]
(again replacing all $|\beta_{j+1}|$ and $|\beta_{j+2}|$ by
$|\beta_{j}|$ and adding the difference to $C_{\beta,m}$). But
this together with the first sum in \eqref{5.4} gives exactly the
first sum in \eqref{5.3}.
\end{proof}

We define
\[
\gamma_j\equiv \beta_{j+2}-2a\beta_{j+1}+\beta_j
\]
then the second, third, and fourth sums in \eqref{5.4} involve
$|\gamma_j|$, $|\gamma_j-\beta_j|$ and $|\gamma_j-\beta_{j+2}|$.
Using $|x-y|^2 \ge |x|^2 + |y|^2 - 2|x||y|$ for the last two, we
obtain (with a new $C_{\beta,m}$)
\begin{align}
(-8)\calI_m \ge &  C_{\beta,m} + \sum O(|\beta_j|^6) + \sum
(2+2|\beta_{j+1}|^2) |\gamma_j|^2 \notag
\\ & + 4 \sum |\beta_{j+1}|^2|\beta_j|^2  -4 \sum |\beta_{j+1}|^2
\big( |\beta_{j+2}| + |\beta_{j}| \big) |\gamma_j| \notag
\\ & + \sum \big|\beta_{j+1}^2-\beta_j^2 \big|^2 - 4a^2
\sum|\beta_{j+1}|^4 \lb{5.7}
\end{align}
since
\[
\log (1-|\beta_j|^2) + |\beta_j|^2 + \tfrac 12 |\beta_j|^4 =
O(|\beta_j|^6)
\]

Next, we use $-4xy\ge -8x^2-\tfrac 12 y^2$ with $x=|\beta_{j+1}|^2
( |\beta_{j+2}| + |\beta_{j}|)$ and $y=|\gamma_j|$ to estimate the
fourth sum by $\sum O(|\beta_j|^6)-\tfrac 12\sum |\gamma_j|^2$.
Also,
\begin{align*}
- 4a^2  \sum |\beta_{j+1}|^4  &=  - \sum|\beta_{j+1}|^2
|\beta_{j+2}+\beta_j-\gamma_j|^2
\\ & \ge  - \sum |\beta_{j+1}|^2
|\beta_{j+2}+\beta_j|^2 - \sum |\beta_{j+1}|^2 |\gamma_j|^2 \\
& \qquad \qquad -2 \sum |\beta_{j+1}|^2|\beta_{j+2}+\beta_j||\gamma_j| \\
& \ge C_{\beta,m} - 4\sum |\beta_{j+1}|^2|\beta_{j}|^2 - \sum |\beta_{j+1}|^2 |\gamma_j|^2 \\
& \qquad \qquad - \sum O(|\beta_j|^6) - \tfrac 14 \sum
|\gamma_j|^2
\end{align*}
again using $-2xy\ge -4x^2-\tfrac 14 y^2$. Plugging these into
\eqref{5.7}, we have
\[
(-8)\calI_m \ge C_{\beta,m} + \sum O(|\beta_j|^6) + \sum \big(
\tfrac 54+|\beta_{j+1}|^2 \big) |\gamma_j|^2 + \sum
\big|\beta_{j+1}^2-\beta_j^2 \big|^2
\]
The last sum is just $\sum \tfrac 12
(|\beta_{j+2}^2-\beta_{j+1}^2|^2 + |\beta_{j+1}^2-\beta_j^2|^2)$
plus a piece that goes into $C_{\beta,m}$. Letting $\veps\equiv
\tfrac 13\min\{2|a|, 2-2|a|\}>0$, we obtain
\begin{align*}
|\beta_{j+1}|^2 |\gamma_j|^2 +\tfrac 12
|\beta_{j+2}^2-\beta_{j+1}^2|^2 + \tfrac 12
|\beta_{j+1}^2-\beta_j^2|^2 \ge \tfrac 12\veps^4|\beta_{j+1}|^4
\end{align*}
Indeed, if the third term is smaller than $\tfrac
12\veps^4|\beta_{j+1}|^4$, then $|\beta_{j}-\beta_{j+1}|$ or
$|\beta_{j}+\beta_{j+1}|$ is less than $\veps|\beta_{j+1}|$, and
similarly for the second term. But then
$|\beta_{j+2}+\beta_j|/|\beta_{j+1}|\in
[0,2\veps)\cup(2-2\veps,2+2\veps)$ and so
$|\gamma_j|/|\beta_{j+1}|\ge \min\{2|a|-2\veps,
2-2\veps-2|a|\}\ge\veps$, meaning that the first term is at least
$\veps^2 |\beta_{j+1}|^4$. So finally,
\[
(-8)\calI_m \ge C_{\beta,m} + \sum O(|\beta_j|^6) +  \sum
|\gamma_j|^2 + \tfrac 12\veps^4\sum |\beta_{j}|^4
\]
that is (by \eqref{5.5} and the definition of $\beta_j$,
$\gamma_j$),
\begin{equation} \lb{5.8}
\calI_m \le C_{\alpha,m} + \sum O(|\alpha_j|^6) - \tfrac 18 \sum
\big| \big\{ (\delta-e^{-i\theta_2})(\delta-e^{-i\theta_1}) \alpha
\big\}_j \big|^2 - \tfrac 1{16}\veps^4\sum |\alpha_{j}|^4
\end{equation}

\begin{proof}[Proof of Theorem \ref{T1.3}]
If the RHS of \eqref{1.8} holds, then the RHS of \eqref{5.4} for
the $n^{\rm th}$ Bernstein-Szeg\H o approximation (with $m\ge n$)
is bounded (in $n$), and so
\[
\inf_n \bigg[ \int \big[1-\cos(\theta-\theta_1)\big]
\big[1-\cos(\theta-\theta_2)\big] \log (w^{(n)}(\theta))
\f{d\theta}{2\pi} \bigg]
> -\infty
\]
By upper semicontinuity of the above integral (which is again an
entropy up to a constant), we obtain the LHS of \eqref{1.8}.

Conversely, assume the LHS of \eqref{1.8} holds. Then the
essential support of $w$ all of $\partial\bbD$, and so by
Rakhmanov's theorem \cite{Rak}, $|\alpha_j|\to 0$. Hence, starting
from some $j$, we have $O(|\alpha_j|^6)\le \tfrac
1{32}\veps^4|\alpha_j|^4$ and so
\begin{equation} \lb{5.9}
\calI_m \le D_{\alpha,m}  - \tfrac 18 \sum \big| \big\{
(\delta-e^{-i\theta_2})(\delta-e^{-i\theta_1}) \alpha \big\}_j
\big|^2 - \tfrac 1{32}\veps^4\sum |\alpha_{j}|^4
\end{equation}
for large $m$ and some bounded (in $m$) $D_{\alpha,m}$. As in the
previous sections, $\int \big[1-\cos(\theta-\theta_1)\big]
\big[1-\cos(\theta-\theta_2)\big] \log (w_m(\theta))
\f{d\theta}{2\pi}$ is bounded above, and so $\calI_m$ is bounded
below by the hypothesis. \eqref{5.9} then shows that the RHS
of \eqref{1.8} holds.
\end{proof}

\bigskip
%%%%%%%%%%%%%%%%%%%%%%%%%%%%%

\end{document}